\documentclass[12pt]{article}

\usepackage{scicite}

\usepackage{times}

\usepackage{graphicx}
\usepackage{amssymb,amsmath}
\usepackage{bm}
\usepackage{dcolumn}
\usepackage{subfigure}
\usepackage{color}

\newenvironment{sciabstract}{%
\begin{quote} \bf}
{\end{quote}}

\newcounter{lastnote}

\title{Deterministic entanglement generation from driving through quantum phase transitions}

\author
{Xin-Yu Luo,$^{1\ast}$ Yi-Quan Zou,$^{1\ast}$ Ling-Na Wu,$^{1\ast}$ Qi Liu,$^{1}$  Ming-Fei Han,$^{1}$\\
Meng Khoon Tey,$^{1,2\dag}$ Li You$^{1,2\dag}$\\
\\
\normalsize $^{1}$State Key Laboratory of Low Dimensional Quantum Physics, \\
\normalsize Department of Physics, Tsinghua University, Beijing 100084, China\\
\normalsize $^{2}$Collaborative Innovation Center of Quantum Matter, Beijing, China\\
\\
\normalsize $^\ast$These authors contributed equally to this work.\\
\normalsize $^\dag$Corresponding author. E-mail: lyou@mail.tsinghua.edu.cn (L.Y.), \\
\normalsize mengkhoon\_tey@mail.tsinghua.edu.cn (M.K.T.).
}

\date{}

\begin{document}


\baselineskip24pt


\maketitle


\begin{sciabstract}
Many-body entanglement is often created through
system evolution, aided by non-linear interactions between the
constituting particles. The very dynamics, however, can also lead to fluctuations
and degradation of the entanglement if the interactions cannot be controlled. Here, we demonstrate near-deterministic generation of an entangled twin-Fock condensate of $\sim11000$ atoms by driving a $^{87}$Rb Bose-Einstein condensate undergoing spin mixing through two consecutive quantum phase transitions (QPTs). We directly observe number squeezing of $10.7\pm0.6$\,dB and normalized collective spin length of $0.99\pm0.01$. Together, these observations allow us to infer an entanglement-enhanced phase sensitivity of $\sim6$\,dB beyond the standard quantum limit and an entanglement breadth of $\sim910$ atoms. Our work highlights the power of generating large-scale useful entanglement by taking advantage of the different entanglement landscapes separated by QPTs.
\end{sciabstract}

Entangled states are fundamental to quantum computation, quantum simulation, and precision measurement.
Their generation constitutes a persistent experimental goal, especially for systems of many particles. Recent breakthroughs
in generating these states demonstrate useful entanglement for beyond standard-quantum-limit precision sensing,
with entangled states generated through spin-twisting~\cite{Ueda1993PhysRevA.47.5138,Treutlein2010NatureSpinSqueezing,Oberthaler2010NatureSpinSqueezing,Strobel2014_Fisher} and spin-mixing dynamics~\cite{Klempt2011ScienceTF,Oberthaler2011NatureSpinmixing,Chapman2011PhysRevLett.107.210406} in condensates,
and through quantum non-demolition (QND) measurements~\cite{KuzmichPhysRevLett.85.1594,Polzik2009QND,Vuletic2010PhysRevLett.104.073604,Cox2016_PRL,Kasevich2016NatureQND} in cold thermal gases. These entangled states are created through
dynamic evolution under nonlinear interactions or QND. They are
invariably not the steady state of the system at any time and sensitively depend
on control and system parameters.

A many-body system can exhibit several quantum phases with different entanglement structures~\cite{FazioNat2002,CramerNatComm2013SpatialEntanglement}.
Tuning the relative strength of competing interactions can induce quantum phase transition (QPT)~\cite{Hansch2002NatureMott} and provides a complementary approach for generating entangled state~\cite{Hansch2002NatureMott,Duan2013PhysRevLett.111.180401}. Here, we demonstrate the power of generating metrologically-useful entanglement by driving a Bose-Einstein condensate (BEC) through QPTs.

Our focus is a BEC of $N$ atoms in a twin-Fock state (TFS), i.e., a fragmented condensate with half of
the atoms ($N/2$) each in two orthogonal modes. This system is deeply
entangled~\cite{Molmer2001PhysRevLett.86.4431,Smerzi2009PhysRevLett.102.100401,Plenio2014PhysRevLett.112.150501,Lewenstein2016random} and enables precision metrology reaching the Heisenberg limit~\cite{Burnett1993PhysRevLett.71.1355}.
Ensembles of TFS have been generated in a number of pioneering experiments~\cite{Chapman2011PhysRevLett.107.210406,Oberthaler2011NatureSpinmixing,Klempt2011ScienceTF}
relying on passive spin-mixing dynamics in atomic BEC~\cite{Zoller2000PhysRevLett.85.3991,Meystre2000PhysRevLett.85.3987}.
Although capable of demonstrating squeezed quantum fluctuations,
TFS samples generated this way exhibit large fluctuations in $N$. In this work, we generate TFS condensate in a nearly deterministic fashion by driving a $^\mathrm{87}$Rb spinor condensate undergoing spin mixing through two consecutive QPTs. Subsequent theoretical considerations reveal that this process is surprisingly robust against system excitations because it is protected by the structure of the system's low-lying eigenstates across the QPTs.

The initial state for our system is a pure $^{87}$Rb condensate in the $m_F=0$ spin component of the $F=1$ ground hyperfine manifold. Its evolution is governed by the following Hamiltonian~\cite{Law1998PhysRevLett.81.5257} (setting $\hbar=1$):
\begin{eqnarray}
H=\frac{c_2}{2N_\mathrm{t}}\left[2(\hat{a}_1^{\dag}\hat{a}_{-1}^{\dag}\hat{a}_0\hat{a}_0+\mathrm{h.c.})+(2\hat{N}_0-1)(N_\mathrm{t}-\hat{N}_0)\right]-q \hat{N}_0,
\label{Ham}
\end{eqnarray}
under the assumption of the same spatial profile for all three spin components~\cite{Law1998PhysRevLett.81.5257}.
The symbols $\hat{a}_{m_F}$ ($\hat{a}_{m_F}^{\dag}$) and $\hat{N}_{m_F}$ denote the annihilation (creation) and the number operators for atoms in the $m_F$ spin component, respectively. The first term in the square bracket describes spin-exchange collisions, through which
correlated atoms in the $m_F=\pm1$ components are created from the $m_F=0$ atoms and vice versa. This spin-mixing process occurs at a rate of $|c_2|$ which is typically a few Hz for $F=1$ $^{87}$Rb condensates ($c_2<0$).
The last term $-q \hat{N}_0$ represents an additional interaction with which the evolution of our system can be manipulated. Here, $q = (\epsilon_{+1}+\epsilon_{-1})/2-\epsilon_0$, with $\epsilon_{m_F}$ being the energy of the $m_F$ component, can be tuned by either external magnetic field or near-resonant microwave dressing field. 
The Hamiltonian in Eq.\,1 conserves the net magnetization $\propto (N_{+1}-N_{-1})$
and the total number of atoms $N_\mathrm{t}=N_{+1}+N_0+N_{-1}$ ($N_{m_F}$ refers to the observed value of $\hat{N}_{m_F}$). Its ground state is determined by the competition
between the $c_2$ and $q$ terms and can take three distinct phases (Fig.\,1A).
For $q \gg 2|c_2|$, the ground state is polar (P phase) with all atoms condensed in the $m_F=0$ component~\cite{Ueda2013RevModPhys.85.1191};
for $q \ll -2|c_2|$, the ground state becomes a TFS (TF phase) with the atoms equally partitioned into the $m_F=\pm1$ components.
The middle region bordered by the two QPT points at $q=\pm 2|c_2|$
corresponds to the broken-axisymmetry (BA) phase whose ground state acquires a transverse magnetization, spontaneously breaking the SO(2) symmetry of the system~\cite{Ueda2013RevModPhys.85.1191,Hoang2016gap}. At the QPT points, the energy gap between the ground state and the first excited state (black solid line in Fig.~1A) becomes smallest, scaling as $7.4 N_\mathrm{t}^{-1/3} |c_2|$~\cite{Duan2013PhysRevLett.111.180401,Hoang2016gap}.

The key idea behind our work is to ramp $q$ down all the way from $q > 2|c_2|$ (P) to $q < -2|c_2|$ (TF) across both QPT points. 
If the ramp were adiabatic, the condensate would stay in the instantaneous ground state. All $N_{\rm t}$ atoms initially in the $m_F=0$ component would be completely converted into the $m_F=\pm 1$ components, giving rise to a TFS with $N_{\rm t}/2$ atoms each in $m_F=\pm 1$ (Fig.~1A).
Because the lifetime of our condensate is typically about 30\,s,
achieving true adiabaticity without significant atom loss is difficult. Nevertheless, our
simulations show that it is still possible to generate TFS samples possessing useful entanglement within the parameter regime implementable in our setup. As a result of small atom loss, the sample generated by our protocol is a mixture of Dicke states with almost balanced populations in the two modes and has properties very similar to those of a TFS.

Our experiments start with $^\mathrm{87}$Rb condensates of $N_\mathrm{t}\sim11800\pm200$ prepared in the $m_F=0$ component with no discernable thermal atoms at a magnetic field of 0.815\,G. After ramping $q$ from $3|c_2|$ to $-3|c_2|$ linearly in $3$\,s,
by tuning the power of the dressing microwave 19-MHz blue detuned from the $F=1$ to $F=2$ hyperfine transition~\cite{ScienceSupplementalMaterial}, the initial condensate is observed to
evolve in an almost deterministic fashion into a TFS sample with a negligible final number of $m_F=0$ atoms (Fig.~1B). To better quantify the evolution process, we define a conversion efficiency, $p_c\equiv N/N_\mathrm{t}=(N_{+1}+N_{-1})/N_\mathrm{t}=1-N_0/N_\mathrm{t}$. Figure~1C shows the behavior of $p_c$ at various instants during the $q$-ramp. During the first 500\,ms of the ramp when $q\ge2|c_2|$, we do not observe any atoms in the $m_F=\pm1$ components. At around 700\,ms, i.e., 200\,ms after passing the first QPT point, $p_c$ starts to grow oscillatorily and reaches $(96\pm2)\%$ in the end. The 200-ms delay is understood to be caused by the system's inability to follow the external drive because of longer system relaxation times near the first QPT point, a manifestation of the Kibble-Zurek dynamics~\cite{Chapman2015kibble}. Modeling the evolution dynamics of Hamiltonian in Eq.\,1 and taking into account the measured atom loss rate~\cite{ScienceSupplementalMaterial}, we reproduce the observed conversion efficiency to excellent agreement without any fitting parameters. In Fig.1C, the black solid line (grey shaded region) represents the theoretical expectation for the mean (standard deviation or s.d.) of $p_c$. The oscillation of the mean $p_c$ is induced by the interference between the populated eigenstates and highlights the many-body coherence of the collective spin-mixing dynamics~\cite{Chapman2011PhysRevLett.107.210406,Oberthaler2011NatureSpinmixing,Klempt2011ScienceTF,Klempt2014PhysRevLett.112.155304}. Within the BA regime, its frequency is about $2|c_2|$ (given by the energy spacings between the low excitations) and its slight damping is caused mainly by the unequal energy spacing between the excited states. The uncertainty of $p_c$ grows gradually across the BA regime, but shrinks considerably (together with the oscillation amplitude of $p_c$) after crossing the second QPT point despite more non-adiabatic excitations.

More in depth understanding of the above observations is gained by theoretically studying the evolution of the excitation spectrum and the distribution of $p_c$ for every excited state. Denoting the $n^\mathrm{th}$ excited eigenstate of Hamiltonian in Eq.\,1 at a given $q/|c_2|$ by $\left|\psi_n\right\rangle  = \sum\nolimits_{k = 0}^{N_{\rm{t}}/2} {{d^n_k}\left| {k,N_{\rm{t}} - 2k,k} \right\rangle }$ with $\left|\cdot\right\rangle$ representing the Fock state $\left| N_{+1},N_0,N_{-1}\right\rangle$, we obtain for each state the average $\overline{p}_{c,n}=\sum\nolimits_k {\frac{2k}{N_{\rm{t}}}|{d^n_k}{|^2}}$ (Fig.~2A) and the standard deviation $\Delta p_{c,n}$ (Fig.~2B). The most notable feature of the $\overline{p}_{c,n}$ and the $\Delta p_{c,n}$ spectra is a $\Lambda$-like structure which reflects the smallest gap positions between the nearest neighbour excited states, and to which the evolution of the system wavefunction $|\Psi(t)\rangle$ is intimately connected. The black solid lines mark the highest excitation $n_\mathrm{max}$, $\sum\nolimits_{n=0}^{n_\mathrm{max}} {|\langle\psi_n(t)|\Psi(t)\rangle|^2}\ge0.99$, created over the 3-s $q$-ramp. The simulated excitation spectra $|\langle\psi_n(t)|\Psi(t)\rangle|^2$ are shown in Fig.~2C. Right after crossing the first QPT point, the system is excited rather appreciably over the lowest $\sim200$ excited states along the canyon to the left of the $\Lambda$-structure with almost negligible spread in its excited states. Subsequently, the excited spectrum spreads out and undergoes oscillations in sync with the average $p_c$ (Fig.\,1C). Upon crossing the ridge to the right of the $\Lambda$-structure (Fig.~2A), the structure of the excitation spectra changes fundamentally.
Overall, the ramp adopted in our experiment creates excitations up to the lowest 14\% of the full energy spectrum, yet it still achieves a $p_c$ of ($96\pm2$)\% and thereby a large entanglement.
This is because the low-lying excited states in the TF phase concentrate narrowly in the high-$N$ Hilbert space, as is shown in Figs.~2A and 2B, by the near unity $\bar{p}_{c,n}$ and the small $\Delta p_{c,n}$ for these states when $q\le-2|c_2|$.

To compare our results with those of previous experiments based on passive spin-mixing dynamics~\cite{Chapman2011PhysRevLett.107.210406,Oberthaler2011NatureSpinmixing,Klempt2011ScienceTF,Klempt2014PhysRevLett.112.155304},
which exclusively work
in the BA regime ($|q|\le 2|c_2|$), we jump to and then stay at $q=0.3|c_2|$ right after preparing $m_F=0$ condensates (at $q=2.2|c_2|$), and measure $p_c$ 500\,ms afterwards. This waiting allows the average $p_c$ to evolve to the maximum, a typical criterion adopted in the earlier experiments~\cite{Chapman2011PhysRevLett.107.210406,Oberthaler2011NatureSpinmixing,Klempt2011ScienceTF,Klempt2014PhysRevLett.112.155304}. Figure~3 contrasts the results of $p_c$ obtained using our `ramped-$q$' method and the usual `fixed-$q$' method (time dependences of the ramps are plotted in the inset). The former approach results in a $p_c=(96\pm2$)\% while the latter gives a broad and, roughly speaking, evenly distributed $p_c$ from 6\% to 88\%. This broad distribution results from the sudden jump over the first QPT point from the initial P phase~\cite{Chapman2011PhysRevLett.107.210406,Oberthaler2011NatureSpinmixing,Klempt2011ScienceTF,Klempt2014PhysRevLett.112.155304}. This jump populates the excited eigenstates of Hamiltonian in Eq.\,1 at the final $q$, which are concentrated within the red dashed line shown in Fig.~2B along the left canyon of the $\Lambda$-structure.  For fixed $q$, the system remains stationary in the corresponding eigenstates (as highlighted by the open circles in Figs.~2A and 2B, for our current example). Despite the small spread in the eigen-energy basis, the large $\Delta p_{c,n}$ of the high-lying excited states in the BA phase (Fig.~2B) explains the broad $p_c$ distribution observed in the `fixed-$q$' approach.

Below, we characterize the qualities of the TFS samples generated using our approach. To prepare TFS samples with smaller atom loss and thus larger quantum-noise squeezing, we adopt a non-linear 1.5-s ramp of $q$ from $2.2|c_2|$ to $-2.2|c_2|$ to obtain the results in Fig.~3 and hereafter (see inset of Fig.~3). The asymmetrical ramp is simulation motivated and experimentally optimized with the aims of minimizing atom loss in the $m_F=\pm1$ components while maintaining a sufficiently small spread in $p_c$. 
With it, we reduce the $m_F=\pm1$ atom loss from about 5\% during the 3-s linear ramp to about 2\%.

As illustrated by the inset of Fig.~4A, an ideal TFS is represented
by a ring along the
equator of the generalized Bloch sphere with vanishing width $\Delta J_z$ and a radius $J_\mathrm{eff}=\sqrt{J_x^2+J_y^2+J_z^2}=J_\mathrm{max}=\sqrt{N/2(N/2+1)}$, where $J_i$ is the $i$-th component of the collective spin \textbf{J}~\cite{Ueda1993PhysRevA.47.5138}.
Figure\,~4A shows the distribution of measured $J_z=(N_{+1}-N_{-1})/2$ for the TFS samples we generate (in blue histogram bars),
based on all 426 data sets from a continuous experiment run over more than
5 hours. Instead of a singular peak with unit probability at $J_z=0$,
the measured distribution has a finite $\Delta J_z$, which gives a number squeezing of $\xi^2=\frac{(\Delta J_z)^2}{N/4}=-10.7\pm0.3$~dB with respect to quantum shot noise (QSN) of a coherent spin state, $\sqrt{N}/2$ (as shown by the red bars) ($N\approx10800\pm400$ atoms for this set of experiment)~\cite{ScienceSupplementalMaterial}. After subtracting the detection noise of $\Delta J_z^\mathrm{DN}\approx10.1$, we infer a number squeezing of $-13.3\pm0.6$\,dB below the QSN. This value compares favorably with previous efforts~\cite{Chapman2011PhysRevLett.107.210406,Oberthaler2011NatureSpinmixing,Klempt2011ScienceTF,Klempt2014PhysRevLett.112.155304} that require heavy post-selections.

The effective length of the collective spin, $J_{\rm{eff}}$, serves as a useful quantitative measure of spin coherence. We determine $J_\mathrm{eff}$ from performing $J_z$ measurement on the TFS rotated by a $\pi/2$ pulse, which turns the TFS from wrapped around the equator into the vertical annulus (inset of Fig.~4B)~\cite{Klempt2014PhysRevLett.112.155304,ScienceSupplementalMaterial}.
Figure~4B shows the histogram of the measured $J_z$ normalized to
$J_{\rm{max}}$ over 1120 continuous runs. The TFS samples we generate exhibit near perfect spin coherence, reflected by a normalized spin length of $\sqrt{\langle J_{\rm eff}^2 /J^2_{\rm max}\rangle}=0.99\pm0.01$ and an almost perfect match of our measured $J_z$-distribution to the black solid line of an ideal TFS. Although the $q$-ramps non-adiabatically populate the approximately few hundred lowest excited states, these states  all exhibit near maximum spin length and zero spin projection $J_z$, thus allowing us to prepare a highly entangled state with high efficiency. Following~\cite{Klempt2014PhysRevLett.112.155304,Molmer2001PhysRevLett.86.4431}, by using the detection-noise subtracted $\xi^2$ and the measured $\langle J_{\rm eff}^2/J_{\rm max}^2 \rangle $, we infer the entanglement breadth of our TFS samples to be at least $910^{+9900}_{-460}$ atoms~\cite{Klempt2014PhysRevLett.112.155304, mcconnell2015entanglement}, or more than 450 atoms at the confidence level of 1 s.d. (see Fig. 4C).

Together, the directly measured number squeezing and the normalized spin length allow us to infer an entanglement-enhanced phase sensitivity of about 6\,dB beyond the standard quantum limit~\cite{ScienceSupplementalMaterial}. Further improvements can come from enhanced atom detection schemes with increased number resolution, or from novel nonlinear detection schemes that can amplify signal strength~\cite{Linnemann2016SU11}.

\noindent {\bf Acknowledgement:} We acknowledge helpful discussions with R. Wang and B. Gao. This work is supported by National Basic Research Program of China (973 program) (grants 2013CB922004 and 2014CB921403), and by National Science Foundation of China (grants 91121005, 91421305, 11574177, 11374176, 11404184, 11654001, and 91636213), and by the Tsinghua University Initiative Scientific Research Program (grant 20111081008).

\noindent {\bf Supplementary Materials:}\\
Materials and Methods\\
Supplementary Text\\
Figures S1-S4\\
References (32-37)\\
\\

\begin{figure}
\centering
\includegraphics[width=0.85\linewidth]{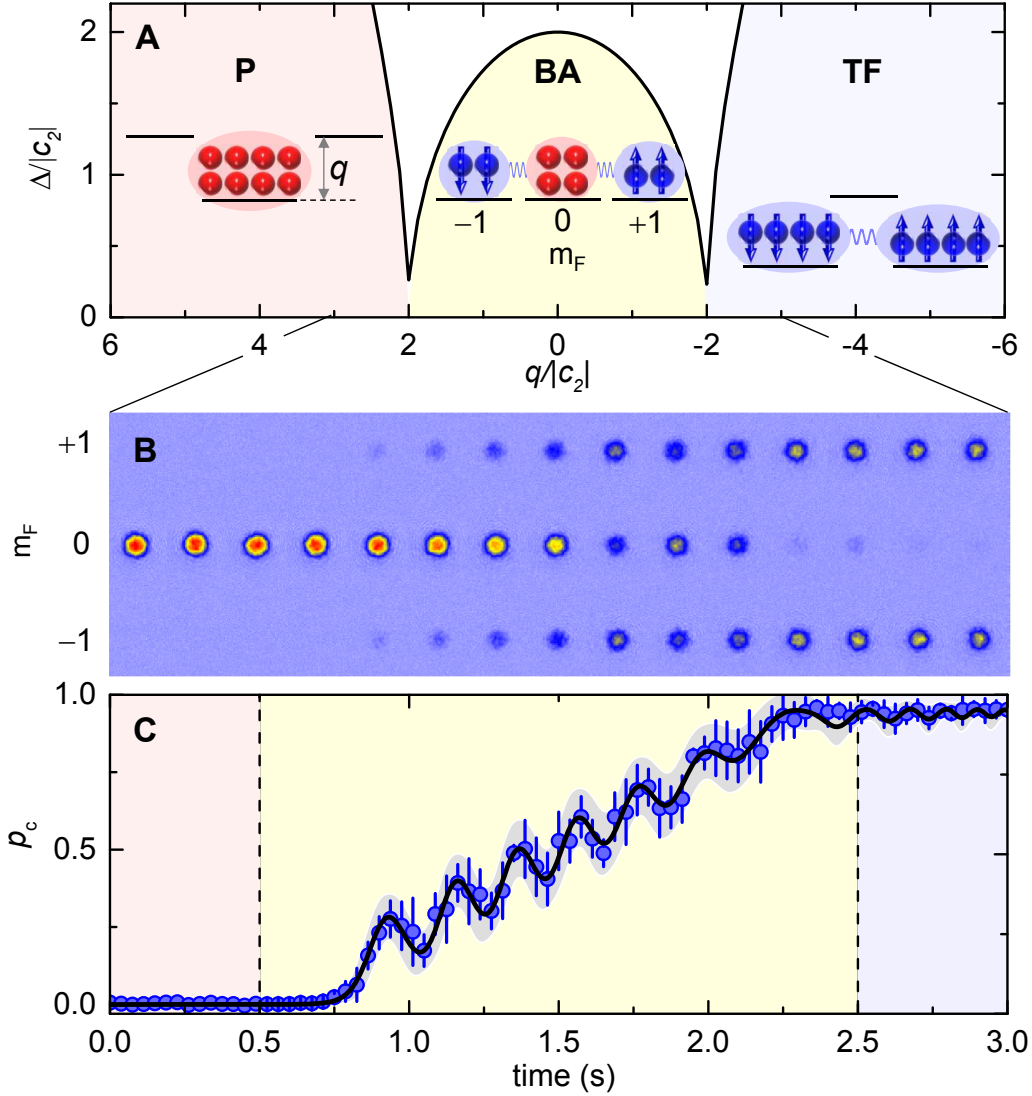}
\caption{
{\bf Efficient generation of twin-Fock state.} (\textbf{A}) The thick black solid line denotes the gap $\Delta$ between the first excited and the ground state
of Hamiltonian in Eq.\,1, which together with the two minima at $q=\pm 2|c_2|$ defines
three quantum phases illustrated by their atom distributions in the three spin components. The first order Zeeman shifts are not shown here because they are inconsequential for a system with zero magnetization.
(\textbf{B}) Absorption images of atoms in the three spin components after Stern-Gerlach separation, showing efficient conversion of a condensate from a polar state into a TFS by sweeping $q$ linearly from $3|c_2|$ to $-3|c_2|$ in $3$\,s.
(\textbf{C}) Conversion efficiency $p_c$ as a function of time. The blue dots mark the experimental results
averaged over 7 runs for each point.
The black solid line (grey shaded region) denotes the theoretical predictions for the mean (standard deviation) of $p_c$ without fitting parameters. The left (right) vertical dashed lines denotes the QPT point at $q=2|c_2|$ ($q=-2|c_2|$).
For all the figures in this work, error bars indicate 68\% statistical confidence interval (1 s.d.).}
 \label{BasicIdea}
\end{figure}

\begin{figure}
\centering
\includegraphics[width=0.85\linewidth]{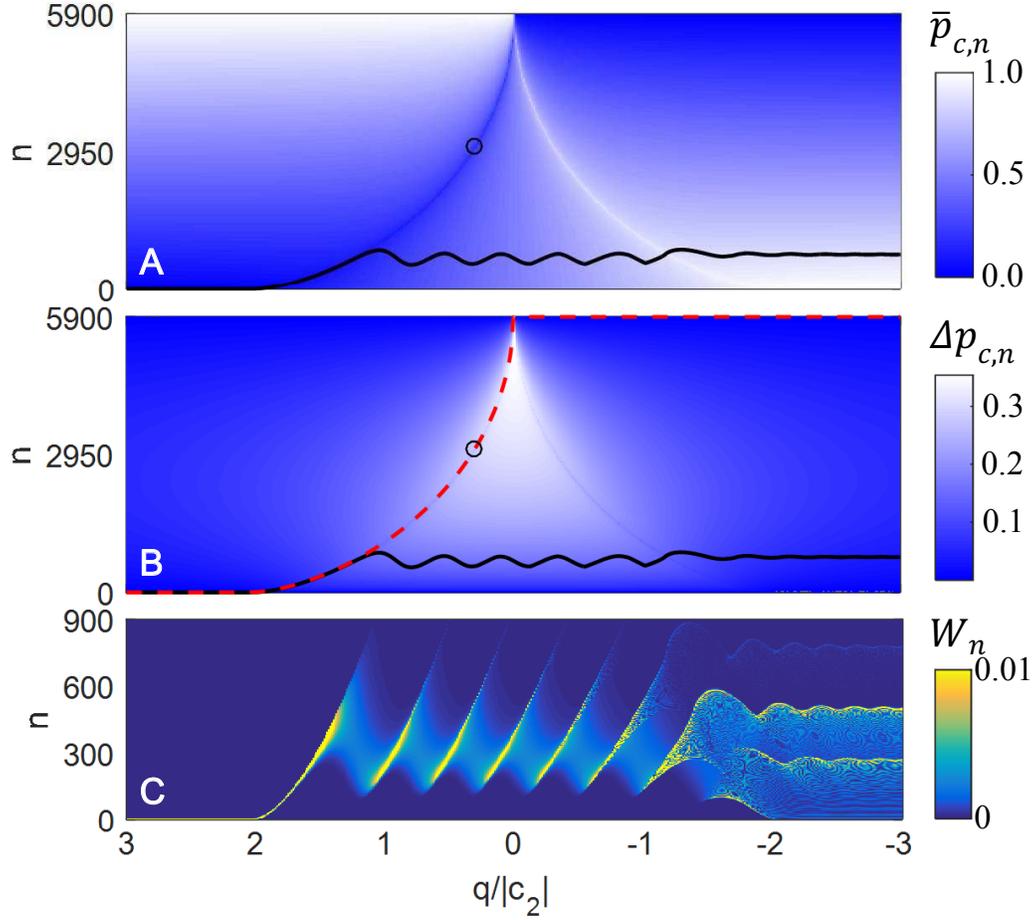}
\caption{
{\bf Simulated conversion efficiency.} Simulated (\textbf{A}) average and (\textbf{B}) standard deviation of $p_c$ of the $n^\mathrm{th}$ excited state at different values of $q/|c_2|$ for a condensate with $N_\mathrm{t}=11800$ atoms. The black solid lines mark the highest excitation for the 3-s $q$-ramp. The red dashed lines and black open circles highlight the excitation spectra for the `fixed-$q$' method (see text). (\textbf{C}) Simulated evolution of the excitation spectra $W_n=|\langle\psi_n(t)|\Psi(t)\rangle|^2$ over the 3-s $q$-ramp.}

\end{figure}

\begin{figure}
\centering
\includegraphics[width=0.85\linewidth]{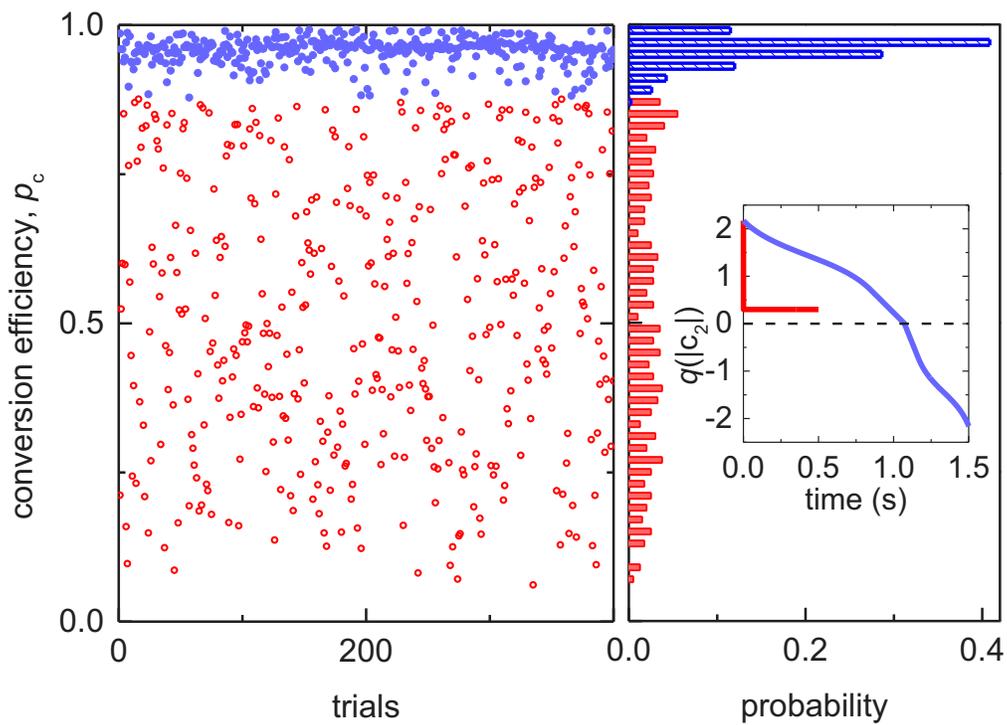}
\caption{
{\bf Comparison to passive scheme.}
The left panel plots the conversion efficiency $p_c$ for the `ramped-$q$' (solid blue) and the `fixed-$q$' (open red circles) methods, with each repeated for 400 trials. The right panel displays the corresponding histograms. The inset shows the time dependence of $q$ for both approaches.}
\end{figure}

\begin{figure}
\centering
\includegraphics[width=0.75\linewidth]{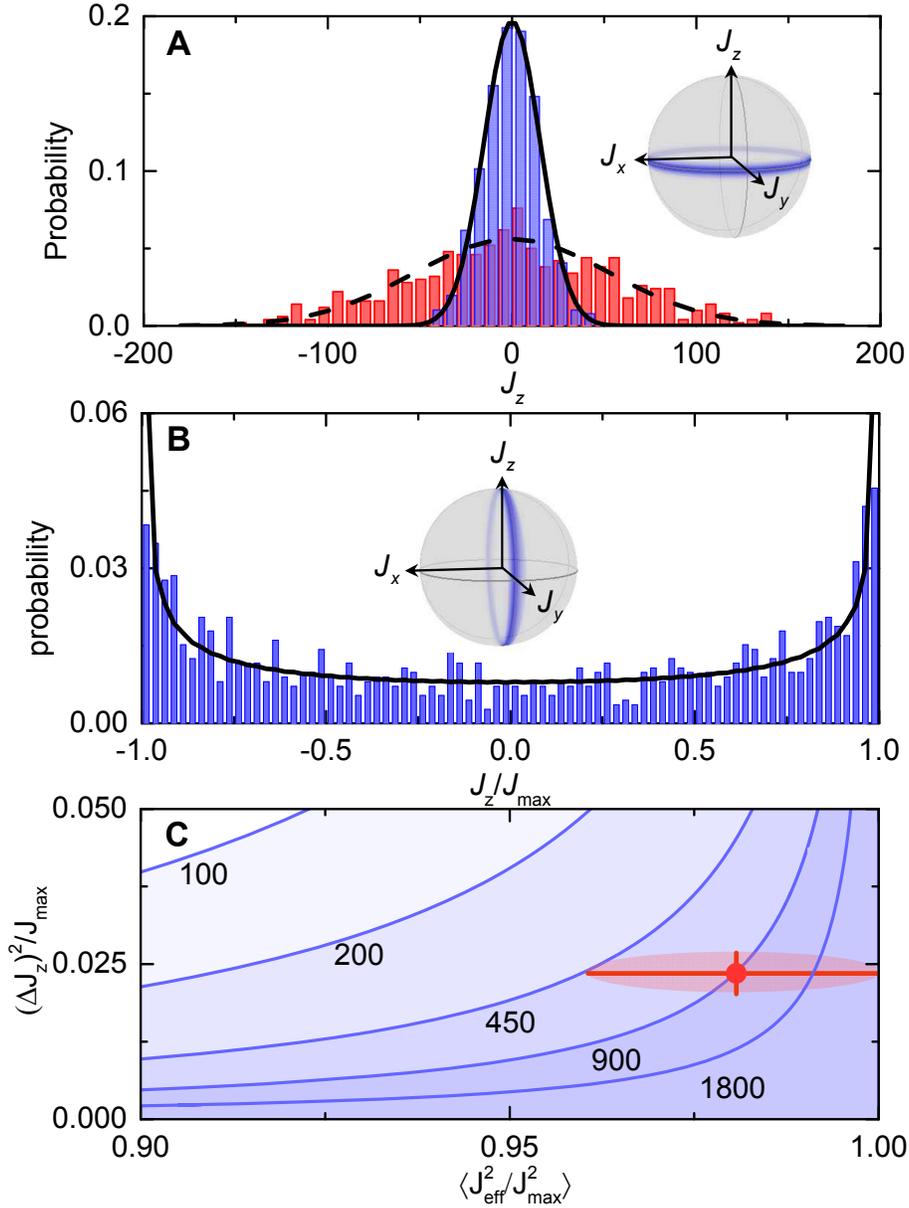}
\caption{
{\bf Characterization of twin-Fock state.} (\textbf{A}) The histogram of the measured $J_z$ for
the generated TFS samples over 426 runs (blue bars) and that for the coherent state measured over 500 samples (red bars).
The solid line denotes the Gaussian fit for the TFS results, whereas the dashed line is the expected envelope for a coherent state with the same number of atoms.
(\textbf{B}) Distribution of $J_z/J_{\rm max}$ for 1120 TFS samples after a $\pi/2$-pulse rotation along the $y$-axis.
The solid line denotes the theoretical distribution of $J_z/J_{\rm max}$ for an ideal TFS with $N$ atoms. Insets in (\textbf{A}) and (\textbf{B}) show the representations of the unrotated and rotated TFSs on Bloch spheres, respectively.
(\textbf{C}) Analysis of entanglement breadth for the TFS samples following refs. \cite{Klempt2014PhysRevLett.112.155304,Molmer2001PhysRevLett.86.4431}. A state below a boundary labeled with number $k$ contains at least a subgroup of non-separable $k$-particle quantum state. The red ellipse represents uncertainties of the measurements at 68\% statistical confidence interval.}
\end{figure}


\begin{thebibliography}{10}
\expandafter\ifx\csname urlstyle\endcsname\relax
  \providecommand{\doi}[1]{doi:\discretionary{}{}{}#1}\else
  \providecommand{\doi}{doi:\discretionary{}{}{}\begingroup
  \urlstyle{rm}\Url}\fi
\providecommand{\bibinfo}[2]{#2}

\bibitem{Ueda1993PhysRevA.47.5138}
\bibinfo{author}{M.~Kitagawa}, \bibinfo{author}{M.~Ueda}.
\newblock \bibinfo{title}{Squeezed spin states}.
\newblock \emph{\bibinfo{journal}{Phys. Rev. A}} \textbf{\bibinfo{volume}{47}},
  \bibinfo{pages}{5138--5143} (\bibinfo{year}{1993}).
\newblock \doi{10.1103/PhysRevA.47.5138}

\bibitem{Treutlein2010NatureSpinSqueezing}
\bibinfo{author}{M.~F. Riedel}, \bibinfo{author}{P.~B{\"o}hi},
  \bibinfo{author}{Y.~Li}, \bibinfo{author}{T.~W. H{\"a}nsch},
  \bibinfo{author}{A.~Sinatra}, \bibinfo{author}{P.~Treutlein}.
\newblock \bibinfo{title}{Atom-chip-based generation of entanglement for
  quantum metrology}.
\newblock \emph{\bibinfo{journal}{Nature}} \textbf{\bibinfo{volume}{464}},
  \bibinfo{pages}{1170--1173} (\bibinfo{year}{2010}).
\newblock \doi{10.1038/nature08988}

\bibitem{Oberthaler2010NatureSpinSqueezing}
\bibinfo{author}{C.~Gross}, \bibinfo{author}{T.~Zibold},
  \bibinfo{author}{E.~Nicklas}, \bibinfo{author}{J.~Est\`eve},
  \bibinfo{author}{M.~K. Oberthaler}.
\newblock \bibinfo{title}{Nonlinear atom interferometer surpasses classical
  precision limit}.
\newblock \emph{\bibinfo{journal}{Nature}} \textbf{\bibinfo{volume}{464}},
  \bibinfo{pages}{1165--1169} (\bibinfo{year}{2010}).
\newblock \doi{10.1038/nature08919}

\bibitem{Strobel2014_Fisher}
\bibinfo{author}{H.~Strobel}, \bibinfo{author}{W.~Muessel},
  \bibinfo{author}{D.~Linnemann}, \bibinfo{author}{T.~Zibold},
  \bibinfo{author}{D.~B. Hume}, \bibinfo{author}{L.~Pezz{\`e}},
  \bibinfo{author}{A.~Smerzi}, \bibinfo{author}{M.~K. Oberthaler}.
\newblock \bibinfo{title}{Fisher information and entanglement of non-Gaussian
  spin states}.
\newblock \emph{\bibinfo{journal}{Science}} \textbf{\bibinfo{volume}{345}},
  \bibinfo{pages}{424--427} (\bibinfo{year}{2014}).
\newblock \doi{10.1126/science.1250147}

\bibitem{Klempt2011ScienceTF}
\bibinfo{author}{B.~L{\"u}cke}, \bibinfo{author}{M.~Scherer},
  \bibinfo{author}{J.~Kruse}, \bibinfo{author}{L.~Pezz{\'e}},
  \bibinfo{author}{F.~Deuretzbacher}, \bibinfo{author}{P.~Hyllus},
  \bibinfo{author}{O.~Topic}, \bibinfo{author}{J.~Peise},
  \bibinfo{author}{W.~Ertmer}, \bibinfo{author}{J.~Arlt},
  \bibinfo{author}{L.~Santos}, \bibinfo{author}{A.~Smerzi},
  \bibinfo{author}{C.~Klempt}.
\newblock \bibinfo{title}{Twin Matter Waves for Interferometry Beyond the
  Classical Limit}.
\newblock \emph{\bibinfo{journal}{Science}} \textbf{\bibinfo{volume}{334}},
  \bibinfo{pages}{773--776} (\bibinfo{year}{2011}).
\newblock \doi{10.1126/science.1208798}

\bibitem{Oberthaler2011NatureSpinmixing}
\bibinfo{author}{C.~Gross}, \bibinfo{author}{H.~Strobel},
  \bibinfo{author}{E.~Nicklas}, \bibinfo{author}{T.~Zibold},
  \bibinfo{author}{N.~Bar-Gill}, \bibinfo{author}{G.~Kurizki},
  \bibinfo{author}{M.~K. Oberthaler}.
\newblock \bibinfo{title}{Atomic homodyne detection of continuous-variable
  entangled twin-atom states}.
\newblock \emph{\bibinfo{journal}{Nature}} \textbf{\bibinfo{volume}{480}},
  \bibinfo{pages}{219--223} (\bibinfo{year}{2011}).
\newblock \doi{10.1038/nature10654}

\bibitem{Chapman2011PhysRevLett.107.210406}
\bibinfo{author}{E.~M. Bookjans}, \bibinfo{author}{C.~D. Hamley},
  \bibinfo{author}{M.~S. Chapman}.
\newblock \bibinfo{title}{Strong Quantum Spin Correlations Observed in Atomic
  Spin Mixing}.
\newblock \emph{\bibinfo{journal}{Phys. Rev. Lett.}}
  \textbf{\bibinfo{volume}{107}}, \bibinfo{pages}{210406}
  (\bibinfo{year}{2011}).
\newblock \doi{10.1103/PhysRevLett.107.210406}

\bibitem{KuzmichPhysRevLett.85.1594}
\bibinfo{author}{A.~Kuzmich}, \bibinfo{author}{L.~Mandel},
  \bibinfo{author}{N.~P. Bigelow}.
\newblock \bibinfo{title}{Generation of Spin Squeezing via Continuous Quantum
  Nondemolition Measurement}.
\newblock \emph{\bibinfo{journal}{Phys. Rev. Lett.}}
  \textbf{\bibinfo{volume}{85}}, \bibinfo{pages}{1594--1597}
  (\bibinfo{year}{2000}).
\newblock \doi{10.1103/PhysRevLett.85.1594}

\bibitem{Polzik2009QND}
\bibinfo{author}{J.~Appel}, \bibinfo{author}{P.~J. Windpassinger},
  \bibinfo{author}{D.~Oblak}, \bibinfo{author}{U.~B. Hoff},
  \bibinfo{author}{N.~Kj{\ae}rgaard}, \bibinfo{author}{E.~S. Polzik}.
\newblock \bibinfo{title}{Mesoscopic atomic entanglement for precision
  measurements beyond the standard quantum limit}.
\newblock \emph{\bibinfo{journal}{PNAS}} \textbf{\bibinfo{volume}{106}},
  \bibinfo{pages}{10960--10965} (\bibinfo{year}{2009}).
\newblock \doi{10.1073/pnas.0901550106}

\bibitem{Vuletic2010PhysRevLett.104.073604}
\bibinfo{author}{M.~H. Schleier-Smith}, \bibinfo{author}{I.~D. Leroux},
  \bibinfo{author}{V.~Vuleti\ifmmode~\acute{c}\else \'{c}\fi{}}.
\newblock \bibinfo{title}{States of an Ensemble of Two-Level Atoms with Reduced
  Quantum Uncertainty}.
\newblock \emph{\bibinfo{journal}{Phys. Rev. Lett.}}
  \textbf{\bibinfo{volume}{104}}, \bibinfo{pages}{073604}
  (\bibinfo{year}{2010}).
\newblock \doi{10.1103/PhysRevLett.104.073604}

\bibitem{Cox2016_PRL}
\bibinfo{author}{K.~C. Cox}, \bibinfo{author}{G.~P. Greve},
  \bibinfo{author}{J.~M. Weiner}, \bibinfo{author}{J.~K. Thompson}.
\newblock \bibinfo{title}{Deterministic Squeezed States with Collective
  Measurements and Feedback}.
\newblock \emph{\bibinfo{journal}{Phys. Rev. Lett.}}
  \textbf{\bibinfo{volume}{116}}, \bibinfo{pages}{093602}
  (\bibinfo{year}{2016}).
\newblock \doi{10.1103/PhysRevLett.116.093602}

\bibitem{Kasevich2016NatureQND}
\bibinfo{author}{O.~Hosten}, \bibinfo{author}{N.~J. Engelsen},
  \bibinfo{author}{R.~Krishnakumar}, \bibinfo{author}{M.~A. Kasevich}.
\newblock \bibinfo{title}{Measurement noise 100 times lower than the
  quantum-projection limit using entangled atoms}.
\newblock \emph{\bibinfo{journal}{Nature}} \textbf{\bibinfo{volume}{529}},
  \bibinfo{pages}{505--508} (\bibinfo{year}{2016}).
\newblock \doi{10.1038/nature16176}

\bibitem{FazioNat2002}
\bibinfo{author}{A.~{Osterloh}}, \bibinfo{author}{L.~{Amico}},
  \bibinfo{author}{G.~{Falci}}, \bibinfo{author}{R.~{Fazio}}.
\newblock \bibinfo{title}{{Scaling of entanglement close to a quantum phase
  transition}}.
\newblock \emph{\bibinfo{journal}{Nature}} \textbf{\bibinfo{volume}{416}},
  \bibinfo{pages}{608--610} (\bibinfo{year}{2002}).
\newblock \doi{10.1038/416608a}

\bibitem{CramerNatComm2013SpatialEntanglement}
\bibinfo{author}{M.~Cramer}, \bibinfo{author}{A.~Bernard},
  \bibinfo{author}{N.~Fabbri}, \bibinfo{author}{L.~Fallani},
  \bibinfo{author}{C.~Fort}, \bibinfo{author}{S.~Rosi},
  \bibinfo{author}{F.~Caruso}, \bibinfo{author}{M.~Inguscio},
  \bibinfo{author}{M.~B. Plenio}.
\newblock \bibinfo{title}{Spatial entanglement of bosons in optical lattices}.
\newblock \emph{\bibinfo{journal}{Nature Communications}}
  \textbf{\bibinfo{volume}{4}}, \bibinfo{pages}{2161} (\bibinfo{year}{2013}).
\newblock \doi{10.1038/ncomms3161}

\bibitem{Hansch2002NatureMott}
\bibinfo{author}{M.~Greiner}, \bibinfo{author}{O.~Mandel},
  \bibinfo{author}{T.~Esslinger}, \bibinfo{author}{T.~W. H\"ansch},
  \bibinfo{author}{I.~Bloch}.
\newblock \bibinfo{title}{Quantum phase transition from a superfluid to a Mott
  insulator in a gas of ultracold atoms}.
\newblock \emph{\bibinfo{journal}{Nature}} \textbf{\bibinfo{volume}{415}},
  \bibinfo{pages}{39--44} (\bibinfo{year}{2002}).
\newblock \doi{10.1038/415039a}

\bibitem{Duan2013PhysRevLett.111.180401}
\bibinfo{author}{Z.~Zhang}, \bibinfo{author}{L.-M. Duan}.
\newblock \bibinfo{title}{Generation of Massive Entanglement through an
  Adiabatic Quantum Phase Transition in a Spinor Condensate}.
\newblock \emph{\bibinfo{journal}{Phys. Rev. Lett.}}
  \textbf{\bibinfo{volume}{111}}, \bibinfo{pages}{180401}
  (\bibinfo{year}{2013}).
\newblock \doi{10.1103/PhysRevLett.111.180401}

\bibitem{Molmer2001PhysRevLett.86.4431}
\bibinfo{author}{A.~S. S\o{}rensen}, \bibinfo{author}{K.~M\o{}lmer}.
\newblock \bibinfo{title}{Entanglement and Extreme Spin Squeezing}.
\newblock \emph{\bibinfo{journal}{Phys. Rev. Lett.}}
  \textbf{\bibinfo{volume}{86}}, \bibinfo{pages}{4431--4434}
  (\bibinfo{year}{2001}).
\newblock \doi{10.1103/PhysRevLett.86.4431}

\bibitem{Smerzi2009PhysRevLett.102.100401}
\bibinfo{author}{L.~Pezz\'e}, \bibinfo{author}{A.~Smerzi}.
\newblock \bibinfo{title}{Entanglement, Nonlinear Dynamics, and the Heisenberg
  Limit}.
\newblock \emph{\bibinfo{journal}{Phys. Rev. Lett.}}
  \textbf{\bibinfo{volume}{102}}, \bibinfo{pages}{100401}
  (\bibinfo{year}{2009}).
\newblock \doi{10.1103/PhysRevLett.102.100401}

\bibitem{Plenio2014PhysRevLett.112.150501}
\bibinfo{author}{N.~Killoran}, \bibinfo{author}{M.~Cramer},
  \bibinfo{author}{M.~B. Plenio}.
\newblock \bibinfo{title}{Extracting Entanglement from Identical Particles}.
\newblock \emph{\bibinfo{journal}{Phys. Rev. Lett.}}
  \textbf{\bibinfo{volume}{112}}, \bibinfo{pages}{150501}
  (\bibinfo{year}{2014}).
\newblock \doi{10.1103/PhysRevLett.112.150501}

\bibitem{Lewenstein2016random}
\bibinfo{author}{M.~Oszmaniec}, \bibinfo{author}{R.~Augusiak},
  \bibinfo{author}{C.~Gogolin},
  \bibinfo{author}{J.~Ko\l{}ody\ifmmode~\acute{n}\else \'{n}\fi{}ski},
  \bibinfo{author}{A.~Ac\'{\i}n}, \bibinfo{author}{M.~Lewenstein}.
\newblock \bibinfo{title}{Random Bosonic States for Robust Quantum Metrology}.
\newblock \emph{\bibinfo{journal}{Phys. Rev. X}} \textbf{\bibinfo{volume}{6}},
  \bibinfo{pages}{041044} (\bibinfo{year}{2016}).
\newblock \doi{10.1103/PhysRevX.6.041044}

\bibitem{Burnett1993PhysRevLett.71.1355}
\bibinfo{author}{M.~J. Holland}, \bibinfo{author}{K.~Burnett}.
\newblock \bibinfo{title}{Interferometric detection of optical phase shifts at
  the Heisenberg limit}.
\newblock \emph{\bibinfo{journal}{Phys. Rev. Lett.}}
  \textbf{\bibinfo{volume}{71}}, \bibinfo{pages}{1355--1358}
  (\bibinfo{year}{1993}).
\newblock \doi{10.1103/PhysRevLett.71.1355}

\bibitem{Zoller2000PhysRevLett.85.3991}
\bibinfo{author}{L.-M. Duan}, \bibinfo{author}{A.~S\o{}rensen},
  \bibinfo{author}{J.~I. Cirac}, \bibinfo{author}{P.~Zoller}.
\newblock \bibinfo{title}{Squeezing and Entanglement of Atomic Beams}.
\newblock \emph{\bibinfo{journal}{Phys. Rev. Lett.}}
  \textbf{\bibinfo{volume}{85}}, \bibinfo{pages}{3991--3994}
  (\bibinfo{year}{2000}).
\newblock \doi{10.1103/PhysRevLett.85.3991}

\bibitem{Meystre2000PhysRevLett.85.3987}
\bibinfo{author}{H.~Pu}, \bibinfo{author}{P.~Meystre}.
\newblock \bibinfo{title}{Creating Macroscopic Atomic Einstein-Podolsky-Rosen
  States from Bose-Einstein Condensates}.
\newblock \emph{\bibinfo{journal}{Phys. Rev. Lett.}}
  \textbf{\bibinfo{volume}{85}}, \bibinfo{pages}{3987--3990}
  (\bibinfo{year}{2000}).
\newblock \doi{10.1103/PhysRevLett.85.3987}

\bibitem{Law1998PhysRevLett.81.5257}
\bibinfo{author}{C.~K. Law}, \bibinfo{author}{H.~Pu}, \bibinfo{author}{N.~P.
  Bigelow}.
\newblock \bibinfo{title}{Quantum Spins Mixing in Spinor Bose-Einstein
  Condensates}.
\newblock \emph{\bibinfo{journal}{Phys. Rev. Lett.}}
  \textbf{\bibinfo{volume}{81}}, \bibinfo{pages}{5257--5261}
  (\bibinfo{year}{1998}).
\newblock \doi{10.1103/PhysRevLett.81.5257}

\bibitem{Ueda2013RevModPhys.85.1191}
\bibinfo{author}{D.~M. Stamper-Kurn}, \bibinfo{author}{M.~Ueda}.
\newblock \bibinfo{title}{Spinor Bose gases: Symmetries, magnetism, and quantum
  dynamics}.
\newblock \emph{\bibinfo{journal}{Rev. Mod. Phys.}}
  \textbf{\bibinfo{volume}{85}}, \bibinfo{pages}{1191--1244}
  (\bibinfo{year}{2013}).
\newblock \doi{10.1103/RevModPhys.85.1191}

\bibitem{Hoang2016gap}
\bibinfo{author}{T.~M. Hoang}, \bibinfo{author}{H.~M. Bharath},
  \bibinfo{author}{M.~J. Boguslawski}, \bibinfo{author}{M.~Anquez},
  \bibinfo{author}{B.~A. Robbins}, \bibinfo{author}{M.~S. Chapman}.
\newblock \bibinfo{title}{Adiabatic quenches and characterization of amplitude
  excitations in a continuous quantum phase transition}.
\newblock \emph{\bibinfo{journal}{Proceedings of the National Academy of
  Sciences}} \textbf{\bibinfo{volume}{113}}, \bibinfo{pages}{9475--9479}
  (\bibinfo{year}{2016}).
\newblock \doi{10.1073/pnas.1600267113}

\bibitem{ScienceSupplementalMaterial}
\bibinfo{note}{Materials and methods are available as supplementary materials
  at the Science website.}

\bibitem{Chapman2015kibble}
\bibinfo{author}{M.~Anquez}, \bibinfo{author}{B.~A. Robbins},
  \bibinfo{author}{H.~M. Bharath}, \bibinfo{author}{M.~Boguslawski},
  \bibinfo{author}{T.~M. Hoang}, \bibinfo{author}{M.~S. Chapman}.
\newblock \bibinfo{title}{Quantum Kibble-Zurek Mechanism in a Spin-1
  Bose-Einstein Condensate}.
\newblock \emph{\bibinfo{journal}{Phys. Rev. Lett.}}
  \textbf{\bibinfo{volume}{116}}, \bibinfo{pages}{155301}
  (\bibinfo{year}{2016}).
\newblock \doi{10.1103/PhysRevLett.116.155301}

\bibitem{Klempt2014PhysRevLett.112.155304}
\bibinfo{author}{B.~L\"ucke}, \bibinfo{author}{J.~Peise},
  \bibinfo{author}{G.~Vitagliano}, \bibinfo{author}{J.~Arlt},
  \bibinfo{author}{L.~Santos}, \bibinfo{author}{G.~T\'oth},
  \bibinfo{author}{C.~Klempt}.
\newblock \bibinfo{title}{Detecting Multiparticle Entanglement of Dicke
  States}.
\newblock \emph{\bibinfo{journal}{Phys. Rev. Lett.}}
  \textbf{\bibinfo{volume}{112}}, \bibinfo{pages}{155304}
  (\bibinfo{year}{2014}).
\newblock \doi{10.1103/PhysRevLett.112.155304}

\bibitem{mcconnell2015entanglement}
\bibinfo{author}{R.~McConnell}, \bibinfo{author}{H.~Zhang},
  \bibinfo{author}{J.~Hu}, \bibinfo{author}{S.~{\'C}uk},
  \bibinfo{author}{V.~Vuleti{\'c}}.
\newblock \bibinfo{title}{Entanglement with negative Wigner function of almost
  3,000 atoms heralded by one photon}.
\newblock \emph{\bibinfo{journal}{Nature}} \textbf{\bibinfo{volume}{519}},
  \bibinfo{pages}{439--442} (\bibinfo{year}{2015})

\bibitem{Linnemann2016SU11}
\bibinfo{author}{D.~Linnemann}, \bibinfo{author}{H.~Strobel},
  \bibinfo{author}{W.~Muessel}, \bibinfo{author}{J.~Schulz},
  \bibinfo{author}{R.~J. Lewis-Swan}, \bibinfo{author}{K.~V. Kheruntsyan},
  \bibinfo{author}{M.~K. Oberthaler}.
\newblock \bibinfo{title}{Quantum-Enhanced Sensing Based on Time Reversal of
  Nonlinear Dynamics}.
\newblock \emph{\bibinfo{journal}{Phys. Rev. Lett.}}
  \textbf{\bibinfo{volume}{117}}, \bibinfo{pages}{013001}
  (\bibinfo{year}{2016}).
\newblock \doi{10.1103/PhysRevLett.117.013001}

\end{thebibliography}
\end{document}